\documentstyle[11pt]{article}

\begin{document}

\begin{center}
\Large{\bf Black holes with regular horizons
in Maxwell-scalar gravity.}
\end{center}
\vskip 26pt
\begin{center}
 Slava G. Turyshev
\footnote{ On leave from Bogolyubov Institute for
Theoretical Microphysics,

\hskip 8pt Moscow State University, Moscow,
Russia, 119899; \hskip 8pt

\hskip 8pt Electronic address: sgt@zeus.jpl.nasa.gov }

\end{center}

\centerline{\sl Jet Propulsion Laboratory MS 301-230,}
\centerline{\sl California Institute of Technology}
\centerline{\sl 4800 Oak Grove Drive - Pasadena,
CA 91109 - USA}
\vskip 2mm

\begin{abstract}

A class of exact   static spherically symmetric
solutions of the Einstein-Maxwell gravity
coupled to a massless scalar field has
been obtained in harmonic coordinates
of the Minkowski space-time. For each value of the coupling
constant $a$, these solutions are
characterized by a set of three parameters, the physical mass
$\mu_0$, the electric charge $Q_0$ and the
scalar field parameter $k$. We have found that the solutions for
both gravitational and electromagnetic
fields are not only affected by the scalar field, but also the
non-trivial coupling with matter constrains the scalar field itself.
 In particular, we have found that the  constant $k$
generically differs from $\pm 1/2$, falling into the interval
$|k|\in  [0, {1\over2}\sqrt{1+a^2} \hskip 2pt ]$. It
takes these values only for black holes or in the case when a
scalar field $\phi$ is totally decoupled from the matter.
Our results differ from those previously obtained in that the
presence of   arbitrary coupling constant $a$   gives an opportunity
to rule out the non-physical  horizons.
In one of the special cases, the obtained solution corresponds
to a charged dilatonic black hole with  only one horizon $\mu_+$
and hence for the Kaluza-Klein case.
The most remarkable property of this result is that the metric,
the scalar curvature, and both electromagnetic and scalar fields
are all regular on this surface. Moreover, while studying the
dilaton charge, we found that the inclusion of the scalar field in
the theory   result in  a contraction of the horizon.
The behavior of the scalar curvature was analyzed.

PACS number(s): \hskip 2pt 04.20.-q,
\hskip 2pt 04.20.Jb, \hskip 2pt 04.40.Nr, \hskip 2pt 04.70.-s

\end{abstract}
\section{INTRODUCTION.}
\vskip 10pt

Recently considerable interest   has been shown in
the physical processes occuring in the strong gravitational
field regime. However, many modern theoretical models
which include the general relativity as a standard gravity
theory, are faced with the problem of the unavoidable appearence
of    space-time singularities.
It is well known that the classical description, provided by general
relativity, breaks down in a domain where the curvature is large,
and, hence, a  proper understanding of such regions requires new
physics \cite{HM}.
The tensor-scalar theories of gravity, where,  the
usual for general relativity tensor field, coexists together
with one or several long-range scalar fields,
are believed to be the most interesting extension of the
theoretical foundation of modern gravitational theory.
The superstring, many-dimensional Kaluza-Klein, and inflationary
 cosmology  theories have
  revived the interest in so-called "dilaton fields",
{\it i.e.} neutral scalar fields whose background values determine
the strength of the coupling constants in the effective
four-dimensional theory. However, although the scalar field naturally
arises in theory, its existence  leads to a violation of
the strong equivalence principle and  modification of
large-scale gravitational phenomena \cite{DGG}, \cite{BH}.
Moreover, the presence of the scalar field
affects  the equations of motion of the
other matter fields as well. Thus, for example, the solutions to the
Einstein-Maxwell-dilaton system were studied in
\cite{GM}-\cite{S}, where it was shown that the scalar field
 generally destroys the horizons. This causes the singularities
 in a scalar curvature to appear on a finite radii.
It is worth noting that the special attention has been
paid to the charged dilatonic black hole
solution presented in  \cite{GHS}. The analysis
 of this solution has shown  that, in the case of $a=0$,
it reduces  to  Reisner-Nordstr\"om solution.
However, for $a\not= 0$, this result
represents qualitatively different
physics. In particular, this solution has a regular
outer event horizon but, for any non-zero value $a$, the
inner horizon is singular. An interesting analogy of the
 behavior of the  black holes and elementary particles
has been  demonstrated in \cite{HW}. Thus, by
analyzing  the perturbations around the extreme holes, the
authors of this previous article have shown the  existence of an
energy gap in the excitation spectrum of the black hole, which
corresponds to the potential barrier isolating them from the
external world.

In order to resolve the dilatonic
black-hole singularities,   the higher-dimentional
extension $(D\ge4)$ of the general relativity
was considered in \cite{GTT}. It was shown that a
dilatonic black hole  with a dilaton coupling constant
$a=\sqrt{ p/(p+2)}$ might be interpreted as a non-singular,
non-dilatonic, black $p$-brane in $(4+p)$ dimentions.
Moreover, when $p$ is even, the $p$-brane resembles the
extreme Reisner-Nordstr\"om solution in that there is still
a curvature singularities inside the horizon. However, when $p$
 is odd, the solution inside the horizon is isomeric to that for the
outside region, and it is completely non-singular.
 There the special interest is presenting the class of
stationary, spherically
symmetric black hole solutions in Kaluza-Klein  theory with $a=\sqrt{3}$,
which in four dimentions, was discussed and classified in \cite{GW}.
The decay of magnetic fields in this theory and the possible
mechanism of the pair creation of monopoles  was analysed in \cite{FD}.
However, even thought these solutions present   interesting
properties in higher dimentions,  their  geometry  become
singular at the classical level for $D=4$.

The motivation for the present work was to find a stable
dilatonic black hole solution, which would demonstrate  non-singular
properties for all possible interacting regimes in four dimentions.
As we shall see later, the covariant generalization of the
harmonic  gauge   presents the necessary opportunity.
In this paper we will focus our attention on the
simplest extension of the standard matter {\it i.e.}
gravity coupled to interacting\footnote{
The harmonic solution, presented in \cite{S}, was obtained
 in the special case $a=0$ when the  interaction between
the matter fields is absent.} electromagnetic and scalar fields.
The density of the Lagrangian function $L_M$ for
the massless scalar and electromagnetic fields is suggested
by the low-energy limit of the string theory and it has the usual
form:
{}
$$ L_M ={1 \over 16\pi}\sqrt{-g} \Big(  -  R +
2  \nabla_n\phi \nabla^n\phi -
 e^{-2 a \phi}F^2\Big),\eqno(1) $$

\noindent where  $  F_{mn}=\nabla_m A_n - \nabla_n A_m$ is the
tensor of the electromagnetic field\footnote{
 The Plank units $\hbar=c= \gamma =1$
are used throughout the paper and  metric convention is
accepted to be $(+ - - -)$.}.
The symmetries of this Lagrangian are the general covariance
and the gauge
symmetry. Besides this, the expression (1) is invariant
under the global scale transformations, namely:
$\phi'(x) = \phi(x) + \phi_c$ and $A'_m(x)  = e^{a\phi_c} A_m(x)$.
This freedom can be eliminated by specifying the value
 of the scalar field at   infinity.
The constant $a$ in (1)  is a dimensionless, arbitrary
parameter.  To study the dependence of the  solutions on
the strength of interaction between the scalar and
electromagnetic fields, an arbitrary coupling constant
\hskip 2pt $a$ \hskip 2pt was introduced in  \cite{GM}, \cite{GHS}.
The arbitrariness of this constant makes it possible to have
both weak $(a\ll 1)$ and strong $(a\gg1)$ coupling regimes.
For  $a=0$, Eq.(1) becomes the standard
Einstein-Maxwell Lagrangian with an extra free massless scalar field.
In the case $a=1$, it corresponds to the contribution in total
action from the low-energy limit of the superstring theory,
treated to the lowest order in world-sheet and string loop expansion.
Since changing the sign of $a$ is equivalent to changing
the sign of $\phi$, we will consider this theory just for $a\ge0$.

In order to clearly state the new results obtained,
the  paper structured  is as follows. In  section 2
 we will derive the basic system of equations for the
gravitational, scalar and electromagnetic fields. The
degree of arbitrariness   caused
by the  covariant Fock - de Donder  harmonic gauge will
be discussed in section 3. The solutions for generalized
radial coordinate and for the scalar field will be obtained
in sections  4 and  5 respectievly. The general static
 spherically symmetric solution for
interacting scalar and electromagnetic fields
 will be presented in section 6. Section 7
is devoted to analysis of  the obtained solution in
 some special cases.
The structure of the   scalar curvature
will be examined in section 8. The
 final results in parametric form will be presented in
section 9, where  we will summarize and suggest future
directions for studying the
behavior of the these  solutions.

\section{THE EQUATIONS OF MOTION.}

By extremizing the  Lagrangian function  $L_M$  (1)
with respect to the metric $g_{mn}$, the electromagnetic potential
$A_m$ and the scalar field $\phi$,
one will obtain  the corresponding field equations.
Those for the gravitational field  take the
following form:
{}
$$ R_{mn}= 8\pi (T_{mn} -{1\over 2} g_{mn}T), \eqno(2.1) $$

\noindent where the symmetric energy-momentum tensor of
the matter fields $T_{mn}$ is calculated to be:
{}
$$T_{mn}= {1 \over 4\pi }\hskip 2pt\Big(\nabla_m\phi\nabla_n\phi
 - {1\over 2}g_{mn}  \nabla_k\phi\nabla^k\phi\Big) + $$
{}
$$+{e^{-2a\phi}\over 8 \pi}\bigg( -2{F_m}^k F_{nk}+
{1\over 2}g_{mn}F^2\bigg).\eqno(2.2)$$

We will be looking for solutions of these equations
which will admit the covariant generalization of the
Fock - de Donder harmonical gauge \cite{S}, \cite{F}, namely:
{}
$$ {\cal D}_m \sqrt{-g} g^{mn} = 0, \eqno(2.3)$$

\noindent where ${\cal D}_m$  is the covariant derivative with
respect to Minkowskii metric $\gamma_{mn}$:
{}
$$\gamma_{mn} = \hbox{diag}\Big(1,-1,-r^2,
-r^2 \sin^2\theta\Big).$$

The equations of motion of the scalar and electromagnetic
fields corresponding to $L_M$ (1) may be
 written as follows:
{}
$$ \nabla_n\nabla^n\phi - {a\over 2} \hskip 2pt e^{-2 a\phi} F^2 =0,
\eqno(2.4) $$
{}
$$ \nabla_m\Big(e^{-2 a\phi}  F^{mn}\Big)=0. \eqno(2.5) $$

To avoid   confusion, let us note that words ``static
spherically symmetrical'' here imply  that not only the
electromagnetic
field $F_{mn}$, but also the electromagnetic potential $A_m$, is
spherically symmetrical and independent of time:
{}
$$\phi(t,r, \theta, \varphi) = \phi(r), \qquad A(t,r,\theta,
\varphi) = \Big(A_0(r), A_1(r),0,0\Big).$$

\noindent Imposing the same conditions on the scalar and
gravitational fields,  one may write
the general form of the effective metric for the static
 spherical symmetric case as follows:
{}
$$ g_{mn} = \hbox{diag}\Big ( \hskip 1mm u(r),
\hskip 1mm - v(r), \hskip 1mm - w(r),
\hskip 1mm - w(r)\hskip 1mm\sin^2\theta \hskip 1pt \Big).
 \eqno(2.6)$$

Then, finally, taking into account the definitions above,
the system of the gravitational field equations (2.1)
may be re-written as follows:
{}
$$R_{00} = {u''\over 2v} +  {u'\over 2v} \big({w'\over w}
 - {v'\over 2v} -  {u'\over 2u}\big) =
{1\over v}(A_0')^2 e^{-2a\phi}, \eqno(2.7a)$$
{}
$$R_{11} = - {u''\over 2u} - {w''\over w} +
{u'\over 2u}\big({u'\over 2u} + {v'\over 2v}\big) +
{w'\over w} \big({w'\over 2w} + {v'\over 2v}\big) =$$
{}
$$=2(\phi')^2 - {1\over u}(A_0')^2 e^{-2a\phi}, \eqno(2.7b)$$
{}
$$R_{22} = - {w''\over 2v} + {w'\over 2v} \big({v'\over 2v} -
 {u'\over 2u}\big) + 1 =
{w\over uv}(A_0')^2 e^{-2a\phi}.\eqno(2.7c)$$

\noindent The equation for the component $R_{33}$
coincides with the one for  $R_{22}$, and
the other equations become exact equalities.
Finally, the equations for the scalar and
 electromagnetic fields from Eqs.(2.4),(2.5) take the form:
{}
$$\phi'' + \phi'\big({u'\over 2u} - {v'\over 2v} +
{w'\over w}\big) =
{a\over u}(A_0')^2 e^{-2a\phi},\eqno(2.8)$$
{}
$$\big({w\over\sqrt{uv}}  A_0'e^{-2a\phi}\big)'=0. \eqno(2.9)$$

The obtained system of equations (2.7)-(2.9) are all that one needs to
find the general spherically symmetric solution to the
Einstein-Maxwell-scalar system (1). In the next section we will examine
the possibility of simplifying of this system of equations
using the covariant gauge conditions  (2.3).

\section{THE  PARAMETERIZATION OF THE SOLUTION.}

The use of the covariant Fock - de Donder
gauge conditions (2.3) significantly symplifyies the system of
the field equations (2.7)-(2.9). To demonstrate this, let
us make a linear combination of the first and third equations
 from the system (2.7) with the coefficients
 $1/u$ and $-1/w$ respectively. The right hand side of the
 obtained relation becomes zero as the matter fields fall out:
{}
$${u''\over 2u} +  {u'\over 2u} \big({w'\over w} -
{v'\over 2v} -  {u'\over 2u}\big)
+ {w''\over 2w} + {w'\over 2w} \big({u'\over 2u} -
{v'\over 2v}\big) - {v\over u} = 0. \eqno(3.1)$$

\noindent  From the gauge condition Eq.(2.3)
one might get another, pure gravitational equation, namely:
{}
$$\big(\sqrt{{u\over v}}w\big)' = 2r\sqrt{uv}. \eqno(3.2)$$

By defining  new functions $\alpha(r)$  and $\beta(r)$ as:
 $ \alpha= \sqrt{uv}, \hskip 2mm  \beta=w\sqrt{{u/v}}$,
one can get the following system of equations
from the Eqs.(3.1) and (3.2):
{}
$$\beta'=2 r\alpha, \qquad (\alpha'\beta^2/\alpha)'=0.
\eqno(3.3)$$

\noindent The general solution of that system might be
written in a parametric form. Indeed, let us present
 $\alpha(p)$ and $\beta(p)$ in a following way:
{}
$$ \alpha(p) = {A \over r_p},\qquad \beta(p)= A
(p^2-\mu^2)\cdot r_p, \qquad r_p = {dr\over dp},\eqno(3.4)$$

\noindent where $A, p$ and $\mu$ are constants (arbitrary
for the
moment). This substitution will enable us to eliminate the
 functions
$\alpha$ and $\beta$ from both equations (3.3) and, as a
result, we will obtain two equations for the same
function $r(p)$:
{}
$$(p^2-\mu^2)\hskip 2pt r_{pp} +
2 p\hskip 2pt r_p  - 2r = 0, \eqno(3.5a)$$
{}
$$(p^2-\mu^2)^2\hskip 2pt r_{pp} + {B\over A^2} = 0,
\eqno(3.5b)$$

\noindent where $B$ is another arbitrary integrating constant.
 Equations (3.5) are easy to integrate; and the common
solution for both of them may  be presented as follows:
{}
$$r(p) = q\bigg[ p + z_0 \bigg(p \ln {{p-\mu}\over {p+\mu}} +
2\mu \bigg)\bigg],\eqno(3.6)$$

\noindent with  arbitrary integrating constants $q, z_0 $ and
  $B= 4\mu^3 q z_0  A^2$. Then from (3.4), one
may write the general solution for the
 system of Eqs.(3.3)  in the   parametric form as follows:
{}
$$\alpha(p) = A\bigg[ 1 +  z_0 \bigg( \ln {{p-\mu}\over {p+\mu}} +
{2\mu p\over{p^2-\mu^2}}\bigg)\bigg]^{-1}, \eqno(3.7a)$$
{}
$$\beta(p) = A\big(p^2-\mu^2\big)
\bigg[ 1 +  z_0 \bigg( \ln {{p-\mu}\over {p+\mu}} +
{2\mu p\over{p^2-\mu^2}}\bigg)\bigg],\eqno(3.7b)$$
{}
$$r(p) =  q\bigg[ p +  z_0 \bigg( p\hskip 2pt
\ln {{p-\mu}\over {p+\mu}} + 2\mu \bigg)\bigg]. \eqno(3.7c)$$

It was noted in   \cite{S}  that the solution of the problem obtained
for the partial case  $ z_0 =0$  might be easily expanded
to a general case with $ z_0 \ne 0$.
Because of this, we will take $ z_0 =0$ from now on and
will defer reconstructing  a non-zero value
of the constant $ z_0 $ to the final results.
The constants $A$ and $q$ are multipliers which define
 the scale of measurements of
the coordinate. Without losing generality, we may set
 these constants to be equal to unity.

The relations (3.7) enable one to express the variables
 $u$ and $v$ as follows:
{}
$$ u(r) = {1\over v(r)}= {{r^2 - \mu^2}\over w(r)}.\eqno(3.8)$$

\noindent It is worth  noting that the first equation from those above
is the usual Schwartzchild condition \cite{GHS}, however, the second
equation, as we shall see later, will be responsible for  qualitatievly
different physics.  By substituting this result into
 the  system of equations  (2.7)-(2.8),  one obtains
{}
$$\big[ -{w'\over w} (r^2 - \mu^2) + 2 r\big]' =
{2 Q^2\over w} e^{2a\phi},\eqno(3.9a)$$
{}
$$ - {w''\over w} + {1\over 2}\bigg({w'\over w}\bigg)^2 =
 2(\phi')^2,\eqno(3.9b)$$
{}
$$\big[\phi' (r^2 - \mu^2)\big]' =  {aQ^2\over w}
e^{2a\phi}.\eqno(3.9c)$$

\noindent The electric charge $Q$ is the integral of the
Maxwell equations (2.9), which generalizes the Gauss' law
for curved space-time in the  following way:
{}
$$ E = {Q\over w} e^{2a\phi}, \eqno(3.10)$$

\noindent where $E = A'_0$ is the intensity of the
electromagnetic field.

To find the solution for the function $w(r)$,
let us define a new function $f(r)$ as follows:
{}
$$w(r) =  f(r) e^{2a\phi(r)}.\eqno(3.11)$$

\noindent Using this expression, the system of the
 equations (3.9) may be rewritten as:
{}
$$\big[ -{f'\over f} (r^2 - \mu^2) + 2 r\big]'f =
2(1+a^2)Q^2,\eqno(3.12a)$$
{}
$$ a \phi'' + a \phi'{f'\over f} + (1+ a^2) (\phi')^2 =
-{1\over 2}\Big[\big({f'\over f}\Big)' +
{1\over 2}\Big({f'\over f}\Big)^2\Big],\eqno(3.12b)$$
{}
$$\big[\phi' (r^2 - \mu^2)\big]'f =  a Q^2.\eqno(3.12c)$$

Our future strategy will be the following: first, we will
 solve the equation $(3.12a)$ for
the function $f$.  Second, the obtained function $f$ will
be used in the equation $(3.12b)$ (which is
considered here as the equation for determining   the scalar
field $\phi$). Solutions for the
functions $f$ and $\phi$, obtained this way, should satisfy
the equation of motion of the scalar field
which is presented by the equation  $(3.12c).$

\section{THE SOLUTION FOR THE FUNCTION  $f(r)$.}

In order to find $f(r)$,
we will introduce a new function $\nu(r)$ by the
following relation:
{}
$$f(r) = 2(1+a^2)Q^2 \cdot (r^2-\mu^2) \nu^2(r), \eqno(4.1)$$

\noindent Using this function, the equation
$(3.12a)$ can be presented in terms of $\nu(r)$ as:
{}
$$\big[ - (r^2 - \mu^2){2\nu'\over \nu}\big]'
(r^2 - \mu^2)\nu^2 =  1. \eqno(4.2) $$

\noindent After some algebra, one may obtain
two solutions for this equation:
{}
$$\nu_1(r) = \pm {1\over 4\sqrt{2}s\mu h}
\Big[B \Big({r-\mu\over r+\mu}\Big)^{s+h} -
 {1\over B}\Big({r+\mu\over r-\mu}\Big)^{s+h}\Big], \eqno(4.3a)$$
{}
$$\nu_0(r) = b \pm {1\over 2\sqrt{2}\mu}
\ln {r-\mu\over r+\mu} , \eqno(4.3b)$$

 \noindent The  constants $h, B, b$ and $s$ are arbitrary
for the moment and,  in general,
may  have  both real and  imaginary values.
It is easy to see that the result $(4.3b)$
is simply the limiting case of the solution $(4.3a)$ with
parameter $h=0$, and therefore the expression for $\nu_1(r)$
from Eq.$(4.3a)$  is the general solution for the
equation (4.2).
Then the  function $f_1(r)$ may be written from
$(4.1)$ and $(4.3a)$ as follows:
{}
$$f_1(r) = (1+a^2){Q^2\over 16\mu^2s^2h^2}  (r^2-\mu^2)
\times$$
{}
$$\times \Big[B\Big({r-\mu\over r+\mu}\Big)^{s+h} -
{1\over B}\Big({r+ \mu \over r-\mu}\Big)^{s+h} \Big]^2.
\eqno(4.4)$$

\noindent Substituting this expression into Eq.$(3.12a)$,
one can see that function $f_1$ becomes the
solution of this equation if the following condition
is satisfied: \hskip 2mm  $s^2h^2 = (s+h)^2=k^2,$
where $k$  is some new
arbitrary parameter. After this, the general solution for
 the function $f(r)$   may finally be presented by the
expression:
{}
$$f(r) = (1+a^2){Q^2\over 16\mu^2k^2}  (r^2-\mu^2)
\Big[B\Big({r-\mu\over r+\mu}\Big)^k- {1\over B}
\Big({r+ \mu \over r-\mu}\Big)^k \Big]^2. \eqno(4.5a)$$

\noindent In the limit $k=0$, this result  will take the form:
{}
$$f_0(r) =  2(1+a^2)Q^2 (r^2-\mu^2)
\left(b \pm {1\over 2\sqrt{2}\mu} \ln{r-\mu
\over r+\mu}\right)^2. \eqno(4.5b)$$

\section{THE SOLUTION FOR THE SCALAR FIELD $\phi(r)$.}

To find the solution for the function $\phi$ from the
equation $(3.12b)$ we will use the  following substitution:
{}
$$ \phi'(r) = {\xi(r)\over {r^2-\mu^2}}, \eqno(5.1)$$

\noindent where $\xi(r)$ is a new function to be determined.
Then, with the help of the expressions $(4.3a)$ and
$(4.5a)$, the equation Eq.$(3.12b)$ becomes:
{}
$$a(r^2-\mu^2)\big(\xi' + 2 \xi{\nu'_1\over \nu_1}\big) +
 (1+a^2)\xi^2 = \mu^2(1-4k^2).\eqno(5.2)$$

\noindent Let us define a new radial coordinate $z$ as follows:
{}
$$z= {{B^2\rho^2 - 1}\over {B^2\rho^2 + 1}},
 \qquad \rho = \left({{r-\mu}\over{r+\mu}}\right)^k.
 \eqno(5.3)$$

\noindent Then the equation (5.2) may be re-written as:
{}
$$\xi_z (1-z^2) + {2\over z}\xi + {{1+a^2}\over 2\mu ka}
 \xi^2 = {\mu\over 2ka}(1-4k^2).\eqno(5.4)$$

\noindent The general solution of this differential equation
has the following form:
{}
$$\xi(\rho)= {2\mu ka\over {1+a^2}}
\bigg({\delta\over k}\hskip 2pt
{C_0^2\big(B\rho\big)^{2\delta/k}+1
\over C_0^2\big(B\rho\big)^{2\delta/k}  -1} -
{{\big(B\rho\big)^2 + 1}\over {\big(B\rho\big)^2 - 1}}
\hskip 2pt\bigg),\eqno(5.5)$$

\noindent where $C_0$  is arbitrary integrating constant and
$\delta$ is given by the expression:
{}
$$\delta=\pm {1\over 2 a}\sqrt{1+a^2-4k^2}. \eqno(5.6)$$

\noindent This finally gives the following general
solution for the function $\phi(r)$:
{}
$$ \phi(r) =  \phi_0 -{a\over 1 + a^2}
 \ln \Big[ B\Big({r-\mu\over r+\mu}\Big)^k  - {1\over B}
\Big({r+\mu \over r-\mu}\Big)^k \Big]+  $$
{}
$$+ {a\over 1 + a^2}\ln \Big[C_0B^{\delta/k}
\Big({r-\mu\over r+\mu}\Big)^{\delta} -
{1\over C_0B^{\delta/k}}
\Big({r+\mu \over r-\mu}\Big)^{\delta}\Big], \eqno(5.7)$$

\noindent where $\phi_0$ is an arbitrary integrating constant.

\section{THE GENERAL SOLUTION.}

Now we are in a position to obtain the general solution
for the function $w(r)$. By substituting the expression
$(4.7a)$ into Eq.(3.11) and expressing
$e^{2a\phi(r)}$ with the help of the  relation (5.7)
we may write the result for $w(r)$ as follows:
{}
$$ w(r) =  (1 + a^2){ Q^2\over 16 \mu^2 k^2}e^{2a\phi_0}
(r^2 - \mu^2)\times$$
{}
$$\times\Big[B\Big({r-\mu\over r+\mu}\Big)^k  - {1\over B}
\Big({r+\mu \over r-\mu}\Big)^k \Big]^{2\over{1+a^2}}
\times$$
{}
$$\times\Big[C_0B^{\delta/k}\Big({r-\mu\over r+\mu}\Big)^{\delta} -
{1\over C_0B^{\delta/k}}\Big({r+\mu\over r-\mu}\Big)^{\delta}\Big]^
{2a^2\over 1+a^2}, \eqno(6.1)$$

\noindent
where the constants $\mu, Q, k, B, C_0$ and $\phi_0$ are
arbitrary for the moment. In order to limit the number of
arbitrary constants in this  solution, we will impose
two asymptotical conditions on functions $\phi(r)$  and
 $w(r)$, namely:
{}
$$\lim_{r\rightarrow \infty}\phi(r) = \phi_0,
\qquad \lim_{r\rightarrow \infty}w(r) = r^2. \eqno(6.2)$$

\noindent  By applying  the first of the conditions
 (6.2) and with the help of the relation (5.7), we    have:
{}
$$C_0B^{\delta/k} -{1\over C_0B^{\delta/k}} =
B-{1\over B}. \eqno(6.3)$$

\noindent
Making of use the second  condition from Eq.(6.2) and
taking into account the result above,
we will obtain another constraint:
{}
$$ \Big( B - {1\over B}\Big)^{-2} =
\hskip 2mm(1+a^2){Q^2 e^{2a\phi_0} \over 16\mu ^2 k^2}.
  \eqno(6.4) $$

These results  enable  us to  write the general
solution for the function $w(r)$. Thus from the expression  (6.1)
with the help of the relations (6.3) and (6.4) we will obtain
this function in the following form:
{}
 $$ w(r) = (r^2 - \mu^2)\Big[{B^2\over B^2-1}
\Big({r-\mu\over r+\mu}\Big)^k
-{1\over B^2-1}\big({r+\mu\over r-\mu}\big)^k\Big]
^{2\over1+a^2}\times$$
{}
$$\times\Big[{C_0^2 B^{2\delta/k}\over C_0^2 B^
{2\delta/k}-1}\Big({r-\mu\over r+\mu}\Big)^\delta-
{1\over C_0^2 B^{2\delta/k}-1}\big({r+\mu\over r-\mu}
\big)^\delta
 \Big]^{2a^2\over{1+a^2}},  \eqno(6.5)$$

\noindent where the constant  $B$ may be defined
by Eq.(6.4) as below:
{}
$$B = \pm{1\over  A}\Big(1\pm\sqrt{1 + A^2}\hskip 1pt\Big),
\qquad  A^2=   (1 + a^2){Q^2e^{2a\phi_0}\over 4\mu^2k^2}.
 \eqno(6.6)$$

\noindent
Making of use the result for the constants $B$ and $C_0$ given by
Eq. (6.3), one might present the final solution for the
function $\phi(r)$ as well. Thus, from the expression (5.7)
we will have
{}
$$ \phi(r) = \phi_0  -{a\over 1 + a^2}\ln
\Big[{B^2\over B^2-1}\Big({r-\mu\over r+\mu}\Big)^k-
{1\over B^2-1}\Big({r+\mu\over r-\mu}\Big)^k\Big] +$$
{}
$$+{a\over 1 + a^2}\ln \Big[{C_0^2 B^{2\delta/k}\over C_0^2 B^{2\delta/k}-1}
\Big({r-\mu\over r+\mu}\Big)^\delta-
{1\over C_0^2 B^{2\delta/k}-1}\Big({r+\mu\over r-\mu}\Big)^\delta \Big].
 \eqno(6.7) $$

We have obtained a general solution for the system of
equations Eqs.(3.9).
 For arbitrary values of the coupling constant $a$, this
 solution is labeled by five arbitrary parameters
 $\mu, Q, k, \phi_0$ and $C_0$. This number of  arbitrary parameters
contradicts   the ``no-hair'' theorem, which states that no parameters
other than mass, electric charge, and angular momentum may be associated
with a black hole. Moreover, the parameters, entering   the solution,
should correspond to three conserved Noether currents for the fields
involved, which are easy to obtain from the Lagrangian (1).
Having this in mind,  we will analyse the obtained
general solution Eqs.(6.5)-(6.7) in the next section in order to
obtain a physically reasonable condition to limit  the  arbitrariness
associated with these parameters.

\section{THE ANALYSIS OF THE SPECIAL CASES.}

In this section we  analyse the  various special
 cases of the general solution
by setting some of the parameters equal to zero  while the
 others remain unchanged.

(i). $a=0$. In this case the result obtained represents the  solution
for the Einstein-Maxwell system with an extra free scalar field.
The dependence on the constant $C_0$ in Eqs.(6.5)-(6.7) drops out and
the solution
may be labeled by the set of three parameters ($\mu$, $k$  and $Q_0$):
{}
$$ \phi(r)\big|_{a=0} =  \phi_0 \pm{1\over 2}\sqrt{1-4k^2}
\ln{r-\mu\over r+\mu},  \eqno(7.1a) $$
{}
$$ w(r)\big|_{a=0} = {1\over 4}( r^2 - \mu^2)\times$$
{}
$$\times
\Big[\Big(1\pm\sqrt{1+A_0^2}\hskip 2pt\Big)
\Big({r-\mu\over r+\mu}\Big)^k+
 \Big(1\mp\sqrt{1+A_0^2}\hskip 2pt\Big)
\Big({r+\mu\over r-\mu}\Big)^k\Big]^2,  \eqno(7.1b)$$

\noindent
 where the constant $A_0$ is given by the expression for
 $A$ (6.6) with $a=0$
This result corresponds to that obtained in \cite{S}. One may notice that
 the scalar field  is real for $|k|\le1/2$, but becomes
 complex when  $|k|>1/2$.
Using the relations for the metric functions (3.8) and (2.6),
we will obtain the actual form of the interval $ds^2$.
Thus, for example, for  $k=\pm1/2$, one may notice that the scalar
field   vanishes (or it becomes constant $\phi=\phi_0$)
 and the solution  takes the form:
{}
$$ds^2 =  {r^2-\mu^2\over
\Big(r+ \sqrt{\mu^2+Q^2} \hskip 2pt \Big)^2} dt^2 - $$
{}
$$- {\Big(r+  \sqrt{\mu+Q^2 }
\hskip 2pt \Big)^2\over r^2-\mu^2}dr^2 -
\Big(r+ \sqrt{\mu^2+Q^2} \hskip 2pt \Big)^2
d\Omega, \eqno(7.2a)$$

\noindent where $d\Omega=d\theta^2 +\sin^2\theta d\varphi^2$.
The parameter $\mu>0$ is related to the physical mass $\mu_0$ as:
{}
$$\mu =  \sqrt{\mu_0^2 - Q^2}.  \eqno(7.2b)$$

\noindent By re-writing the expression (7.2a) in terms of the
physical mass $\mu_0$, we esstablish the correspondence of this
result to the solution of the Reisner-Nordstr\"om type,
obtained in harmonic coordinates of the Minkowski  space-time,
which may be presented as follows:
{}
$$ds^2 = \Big({r-\mu_0\over r+\mu_0}+
{Q^2\over (r+\mu_0)^2} \Big) dt^2 - $$
{}
$$-\Big({r-\mu_0\over r+\mu_0}+
 {Q^2\over (r+\mu_0)^2}\Big)^{-1}dr^2 -
\big(\hskip 1pt r+\mu_0\hskip 1pt\big)^2 d\Omega. \eqno(7.2c)$$

\noindent This result, unlike
the usual Reisner-Nordstr\"om solution, has one horizon
only, which is given by the expression (7.2b). This
seems to be quite reasonable. Indeed, from the beginning we have
been looking for a solution outside the source
where the energy-momentum tensor of  matter distribution
equals   zero. As a result, one may show that the physical
radius of the horizon $r_0(\mu)$ is positive and always
remains outside  the body's mass shell $r_0(\mu)\ge\mu_0$,
which is unlike  the conventional Reisner-Nordstr\"om solution,
where the ``inner'' horizon is always inside the mass surface
$r_-\le\mu_0$ and equal to it in the extreme case only.

(ii). An interesting case is arises in  the strong
interaction regime  when $a\gg1$. Examining the expressions
(6.5)-(6.6) in the extreme regime of $a\to\infty$ and $k=\pm1/2$ one
obtains $\phi(r)= \phi_0$ and the following expression for $w(r)$:
{}
$$ w(r) = ( r^2 - \mu^2)\Big[{C_0^2 \over C_0^2-1}
\Big({r-\mu\over r+\mu}\Big)^{\pm{1/2}}-
{1\over C_0^2 -1}\Big({r+\mu\over r-\mu}\Big)^
{\pm{1/2}} \Big]^2.  \eqno(7.3a)$$

This solution corresponds to that of the
Reisner-Nordstr\"om type with the
``induced charge'' $J$ generated by the constant $C_0$.
Taking,  for example, the minus sign in the
 powers of expression in $(7.3a)$, one obtains the following
expression:
{}
$$ds^2=  \Big({r-\hat{\mu}\over r+\hat{\mu}}+
{J^2\over (r+\hat{\mu})^2} \Big) dt^2 - $$
{}
$$- \Big({r-\hat{\mu}\over r+\hat{\mu}}+
{J^2\over (r+\hat{\mu})^2} \Big)^{-1}dr^2 -
\big(\hskip 1pt r+\hat{\mu}\hskip 1pt\big)^2
d\Omega, \eqno(7.3b)$$

\noindent where parameters $\mu$ and $C_0$ are connected to
 physical mass $\hat{\mu}$ and ``charge''  $J$ as follows:
{}
$$ \mu = \hat{\mu} {C_0^2 -1\over C_0^2 +1}
 \qquad J = {2\hat{\mu} C_0 \over{C_0^2 +1}}.$$

This result is quite surprising.  Indeed, taking
the limit $a\to \infty$ and  the condition $k=\pm1/2$
is equivalent to cutting
off both the electromagnetic and the scalar terms in
the Lagrangian density $L_M$ (1).
Then, because of no matter fields are present,
 this solution should be one for a pure static spherically
 symmetric gravity.
Instead, as a result, one obtains the solution (7.3b) of the
 Reisner-Nordstr\"om type with the effective metric
 similar to that in (7.2c). Since the scalar field is
responsible for appearance of the constant $C_0$, then the
``induceed charge'' $J$ is caused by the scalar field, which
is absent! In order to resolve this apparent paradox,
we must  require that $C_0=0$.
This condition simply corresponds to the
 renormalization of the constant $\phi_0$ in Eq.(6.1).
Then  the expression  (7.3b) becomes the Fock solution
for static spherically symmetric black hole
in harmonic coordinates of the Minkowskii space-time:
{}
$$ds^2 =  \Big({r-\mu\over r+\mu}\Big) dt^2 -
\Big({r+\mu\over r-\mu}\Big) dr^2 -
 \big(r+\mu\big)^2 d\Omega. \eqno(7.4)$$

\noindent This solution has one horizon at
$\mu \Rightarrow\mu_+=\mu_0$.
Note, the physical radius of the horizon  is always positive and
equals  $r_0(\mu)=2\mu_0$.

(iii). $Q= 0$. When the electric charge vanishes, the solution
 reduces to that of pure scalar gravity with
interval $ds^2$ written as:
{}
$$ds^2=  \Big({r-\mu\over r+\mu} \Big)^q dt^2 -
 \Big({r+\mu\over r-\mu} \Big)^q dr^2 -
\big(\hskip 1pt r^2 -\mu^2\hskip 1pt\big)
 \Big({r+\mu\over r-\mu} \Big)^q  d\Omega, \eqno(7.5a)$$
\noindent where the constant $q$ given by the relation
{}
$$ q=2\hskip 2pt{k+a^2\delta\over 1+a^2}=
{1\over 1+a^2}\Big({2k\pm a\sqrt{1+a^2-4k^2}}
\hskip 2pt\Big).\eqno(7.5b)$$

\noindent
The scalar field $\phi(r)$ for this case is represented by
the following expression:
{}
$$\phi(r) = \phi_0+
{1 \over 2( 1+a^2)}\Big( 2ak \pm \sqrt{1+a^2-4k^2}
\Big)\ln\Big({r-\mu\over r+\mu}\Big).  \eqno(7.5c) $$

\noindent The parameter $\mu\ge0$ defines the location of
one horizon $(\mu_+)$ which is, in the case (7.5),
related to the physical mass $\mu_0$ as $\mu_+ =
 \mu_0/q$ and, for any $q\not =1$, this horizon  is
singular\footnote{The  condition $q=1$ may be written
equivalently  as $(2k\pm1)^2(1+a^2)=0$.}.

Note  that taking $Q=0$ is equivalent to dropping the
electromagnetic term from the Lagrangian density
$L_M$ (1). However, one might find  quite unexpectedly
that, even after taking $Q\to 0$, our results
 still depend on the arbitrary
parameter $a$ which characterizes the
 intensity of the interaction between the matter fields.
This contradiction might be resolved
by both taking the parameter $k$ to be $k=\pm1/2$
and by choosing  the signs in (7.5) in such a way that
these expressions will not depend on  $a$.
This suggests that it is not only the scalar field
that affects the solutions for the gravitational and
electromagnetic
fields, but also the interaction between the matter fields
that puts constraints on the scalar field itself.
The usual Fock solution (7.4) might be obtained
from  (7.5) by setting $q=1 $   and choosing
the same signs for both terms in $(7.5b)$.

(iv). One might expect that all the expressions for the general
 solution should omit the homogeneous non-trivial
limit in the case where constant $a$ becomes imaginary:
 $a \to\pm i$.  In that limit one will obtain the following
assymptotically flat (with  $\phi_0=0$) result:
{}
$$\phi(r)\big|_{a\to \pm i} = \mp i  \Big[{1-4k^2\over 8k}
\hskip 2pt \ln{r+\mu\over r-\mu} \hskip 2pt +$$
{}
$$+ {Q^2\over 16\mu^2 k^2}\Big(1- \Big({r-\mu\over r+\mu}\Big)^{2k}
\hskip 2pt \Big) \Big],\eqno(7.6a)$$
{}
$$ w(r)\big|_{a\to \pm i}  = ( r^2 - \mu^2)
\Big({r+\mu\over r-\mu}\Big)^{1+4k^2\over 4k}
\exp \Big[ {Q^2\over 8\mu^2 k^2}
\Big(1- \Big({r-\mu\over r+\mu}\Big)^{2k}
\hskip 2pt \Big) \Big].  \eqno(7.6b)$$

\noindent It is easy to see that these expressions are,
in general, singular.
However, choosing the parameter $k=\pm1/2$,  one might
avoid this singularity.
Thus, for $k=1/2$  the expressions (7.6) become:
{}
$$ \phi(r)= \mp i\hskip 2pt{ Q^2\over 4\mu^2}\Big(1-
{r-\mu\over r+\mu}\hskip 2pt\Big),\eqno(7.7a) $$
{}
$$ w(r) = ( r + \mu)^2 \exp\Big[{Q^2\over 2 \mu^2}
\Big(1- {r-\mu\over r+\mu}\hskip 2pt\Big)\Big], \eqno(7.7b)$$
{}
$$u(r)={1\over v(r)} = \Big({r-\mu\over r+\mu}\Big)
\exp\Big[{Q^2\over 2\mu^2}\Big({r-\mu\over r+\mu}\hskip 2pt-
1\Big)\Big]. \eqno(7.7c)$$

\noindent
This is an interesting generalization of the Fock solution
 (7.4) in the presence of the
electromagnetic and pure complex scalar fields.
 Solution (7.7) is labeled by the means of two
parameters $\mu$ and $Q$.
This  solution has   regular event horizon  $r_+$, which is
related to the physical mass $\mu_0$ and the charge $Q$
of the black hole as follows:
{}
$$r_+ \quad \Rightarrow \quad
\mu_+= {1\over 2}\Big(\mu_0+
\sqrt{\mu_0^2- 2Q^2}\hskip 2pt\Big). \eqno(7.8)$$

\noindent The expression (7.8) limits the possible value of the
 physical mass to be  $\mu_0\ge\sqrt{2}|Q|$.
It is easy to show that the scalar curvature
 corresponding to the solution $(7.7)$ is also regular
on surface  (7.8).

The presence of $i$ in the expression for $\phi$ in
(7.7a) might be interpreted
as changing the sign in front of the scalar field's kinetic
term \cite{T} in the Lagrangian density $L_M$ (1) to be:
{}
$$ L_M = -{1\over 16 \pi}\sqrt{-g}\Big(R +
2 g^{mn}\nabla_m  \varphi \nabla_n \varphi +
  e^{-2\varphi}  g^{mn}g^{kl}F_{mk} F_{nl}\Big), \eqno(7.9)$$

\noindent where we denote $\varphi = -i\phi$.
Unfortunately,  the negative kinetic term
$-g^{mn}\nabla_{m}\varphi\nabla_{n}\varphi$ in (7.9) generally
leads to a theory  without stable ground state and, moreover,
it allows infinitely many negative energy states
when the system is quantized \cite{BH}, \cite{S}.

(v). In the case of $k=0$ the general
solution of Eqs.(6.5)-(6.7)  becomes:
{}
$$ \phi(r)\big|_{k=0} = \phi_0 -{1\over2}{1\over
\sqrt{1 + a^2}}\ln \Big({r-\mu\over r+\mu}\Big)-$$
{}
$$-{a\over 1 + a^2}
 \ln \Big[1 -\sqrt{1 + a^2}{Q_0\over 2\mu}\ln
\Big({r-\mu\over r+\mu}\Big) \Big], \eqno(7.10a)$$
{}
$$ w(r)\big|_{k=0} = ( r^2 - \mu^2)\times$$
{}
$$\times \left[1 -\sqrt{1 + a^2}{Q_0\over 2\mu}
\ln \Big({r-\mu\over r+\mu}\Big) \right]^{2\over{1+a^2}}
 \Big({r+\mu\over r-\mu} \Big)^
{a\over{\sqrt{1+a^2}}}.\eqno(7.10b)$$

With  $Q_0=0$ and an arbitrary $a$, this result represents the
usual solution for the scalar field given by (7.5) with $k=0$.
 For an arbitrary value of both parameters
 $a$ and $Q_0$, the  expressions (7.10)  represent the
solution with the  singularity  at $r=\mu$.
The parameter $\mu$ is related to physical mass $\mu_0$
and electric charge $Q_0=Qe^{a\phi_0}$  as follows:
{}
$$a\mu =\mu_0\sqrt{1+a^2}-Q_0. \eqno(7.10c)$$

\noindent Taking into account that  parameters
$a$ and $\mu$ are both  non-negative $a, \mu>0$,
 one may conclude that this last relation saturates the
bound  $\mu_0\sqrt{1+a^2}\ge Q_0$. If we further
set the parameter $\mu=0$ (extreme case), the result
(7.10) will take the form:
{}
$$ \phi(r)\Big|_{\mu=0\atop k=0} = \phi_0-{a\over 1 + a^2}
 \ln \Big[1 + (1 + a^2){\mu_0\over r} \Big],\eqno(7.11a)$$
{}
$$ w(r)\Big|_{\mu=0\atop k=0} = r^2  \Big[1 +
(1 + a^2){\mu_0\over r}\Big]^{2\over 1+a^2}.\eqno(7.11b)$$

\noindent
It easy to see from  $(7.10c)$, that the physical mass
$\mu_0$ is, in this case, generated by  the electric charge only.
For any $a\not=0$, this expression is singular  at $r=0$, however,
in the case $a=0$ result (7.11)
corresponds to extreme case of the Reisner-Nordstr\"om solution (7.2c).

(vi). And finally, in the case $k=\pm1/2$,  the general solution
 becomes a charged dilatonic black hole solution in harmonic coordinates.
To show this, let us take, for example,
$k=1/2$ and a negative sign in front of
the $\sqrt{1+A^2}$ in (6.6). Then one will obtain the following
result\footnote{Result for the value $k=-1/2$
might be obtained by changing $\mu \rightarrow -\mu$ in
the expressions (12) and   will correspond to
a solution with a negative physical mass.}:
{}
$$\phi(r)=\phi_0+ {a\over1+a^2}
\ln\bigg({r+\mu
\over r+\sqrt{\mu^2+ Q_\star^2}}\bigg), \eqno(7.12a)$$
{}
$$w(r)=\Big(r+ \sqrt{\mu^2+ Q_\star^2}\Big)^2
\bigg( {r+\mu\over r+
\sqrt{\mu^2+ Q_\star^2} }\bigg)^{2a^2\over1+a^2},\eqno(7.12b)$$
{}
$$u(r)={1\over v(r)}=\Big( {r-\mu\over r+
\sqrt{\mu^2+ Q_\star^2}}\Big)\bigg({r+\mu\over r+
\sqrt{\mu^2+ Q_\star^2}}\bigg)^{{1-a^2}\over{1+a^2}},\eqno(7.12c)$$

\noindent where  the parameter  $\mu\ge0$ and the constant
 $Q^2_\star$ is defined by Eq. (6.6) as:
{}
$$  Q^2_\star= (1 + a^2)Q^2e^{2a\phi_0}.\eqno(7.13)$$

\noindent This  solution, unlike the result discussed in
\cite{GHS}, has only one horizon $\mu_+\ge0$,
which is given as follows:
{}
 $$\mu_+= {1\over 1-a^2}\Big(\sqrt{\mu_0^2 -
(1-a^2)Q^2_0}-\mu_0a^2\Big). \eqno(7.14a)$$

\noindent In the case $a=0$ this result corresponds to that of
Eq.(7.2b) and for $a=1$ one will have:
{}
$$\mu_+\Big|_{a=1}= \mu_0-{Q^2_0\over 2\mu_0}.  \eqno(7.14b)$$

\noindent The expression (7.14a) saturates the   bound
for the charge $Q_0=Qe^{a\phi_0}$ of the hole as presented below:
{}
$$ \mu_0^2(1+a^2) \ge Q_0^2 \eqno(7.14c)$$

\noindent Furthermore, one may show that horizon $\mu_+$ is
the regular function of the coupling parameter $a$, so that the
function $\mu_+(a)$ remains non-negativite $\mu_+\ge0$ and
finite $\mu_+\le {\mu_+}_{max}<\infty$ for all values of $a$:
$a \in [0,\infty \hskip 2pt[$.  The physical radius  of the horizon
$r_0(\mu)$ is defined by the expression $r^2_0(\mu)  = w(r)_{r=\mu}$,
which in terms of the physical mass $\mu_0$ and charge $Q_0$
gives the following result:
{}
$$r^2_0(\mu)  =\left(\mu_0+
\sqrt{\mu_0^2-(1-a^2)Q_0^2}\right)^{2\over 1+a^2}\times$$
{}
$$\times\Bigg( {2\over 1-a^2}\Big[\sqrt{\mu_0^2-(1-a^2)Q_0^2}-
2\mu_0 a^2\Big]\Bigg)^{2a^2\over1+a^2}. \eqno(7.15a)$$

\noindent Thus, in the case $a=0$, the radius $r_0$
is given as follows:
{}
$$r_0(\mu)\Big|_{a=0}=  \mu_0+\sqrt{\mu_0^2-Q^2}. \eqno(7.15b)$$

\noindent This expression always remains finite
and reaches its minimum value $\mu_0$ for the extreme case,
when $\mu_0=|Q|$. When $a=1$, the expression (7.15a) behaves as:
{}
$$r_0(\mu)\Big|_{a=1}=  2\mu_0\left(1-{Q_0^2\over2\mu^2_0}
\right)^{1/2}. \eqno(7.15c)$$

\noindent Note, that for any $a>0$ the   radius $r_0(\mu)$ vanishes
for the extreme hole, and the  geometry  becomes singular.

Concluding this section we would like to emphasise  that
in order for the general solution \big(Eqs.(6.5) and (6.7)\big) to
correspond to a black hole, it may contain only  three parameters,
namely: $\mu, a$ and $Q e^{a\phi_0}$. This conclusion
is in accord with the ``no-hair'' theorem and, in   the
static spherically symmetric case, instead of the angular
momentum,   the coupling constant $a$  may be added to the set
of the possible parameters. This result may be interpreted as if
the conserved dilaton charge $D$ becomes an additional parameter of
the solution. The dilaton charge is defined by the statement
that at infinity $\phi(r)\rightarrow \phi_0+D/r+{\cal O}(r^{-2})$,
which for the   case (7.12) gives the   following result:
{}
$$D={a\over 1-a^2}\Big(\sqrt{\mu_0^2-(1-a^2)Q_0^2}-\mu_0\Big).\eqno(7.16)$$

\noindent
One can see that this expression is always negative.
It vanishes in the case $a=0$, for $a=1$ it becomes $D=-Q_0^2/(2\mu_0)$,
and it  approaches the assymphotic value, $D_\infty=-|Q_0|$,
when $a\rightarrow \infty$. Note, that $D$ is not an independent parameter,
however, if one decides to account for it as being independent, one will
obtain an interesting modification of the expression  (7.14a) for
the horizon of the solution:
{}
$$\mu_+=\mu_0-{|D|\over a}.\eqno(7.17)$$

\noindent This result means that inclusion
of the scalar field in the theory is leading to a contraction of the
horizon. The usual condition $\mu\ge0$ saturates the
boundary for the reduced dilaton charge
$\hat{D}_a=|D|/a$  as $\mu_0\ge\hat{D}_a$.
The horizon vanishes for the extreme case, when  $\hat{D}_a=\mu_0$,
leading to a naked singularity.

It is easy to verify that  in the
limit $Q\to 0$, the expressions  $(7.12)$ correspond  to the
 Fock solution (7.4) independent of the value for the constant
$a$.  Indeed, taking $Q=0$ is equivalent
to extracting the electromagnetic term from the
action (1). Moreover, by choosing the  parameter $k$
to be   $k=\pm1/2$, we will also eliminate
 the term corresponding to a free scalar field.
Then, the solution in this limit should describe a static
spherically symmetric distribution
of matter. This analysis suggests that
the result   $(7.12)$  corresponds to a  charged
dilatonic black hole solution in harmonic coordinates.
In the next section we will show that not only
the metric, the electromagnetic field and the scalar field
of the  solution  $(7.12)$ are regular on the surface  $(7.14a)$,
but, in addition to this, the scalar curvature $R(r)$
is also remains finite.

\section{THE SCALAR CURVATURE.}

It is well known that the simplest way to study the behavior
of the scalar curvature $R$ is to use the  gravitational field
equations. Indeed, as long as the electromagnetic
part of the energy-momentum tensor Eq.(2.2) is traceless,
the only contribution to the curvature $R$ comes from the
scalar field $\phi$. Thus, by taking the trace of the
Hilbert-Einstein equations (2.1), one can present the
scalar curvature $R$ as:
{}
$$R(r) = -8\pi T = 2g^{mn}\nabla_m\phi\nabla_n\phi =
 -{2\phi'^2(r)\over w(r)} (r^2-\mu^2).\eqno(8.1)$$

\noindent
Substituting  the results for $\phi(r)$ and $w(r)$ from
the general solution presented by the Eqs. (6.5) and (6.7)
in the expression above, we will obtain the expression for
the  corresponding scalar curvature $R$
as follows:
{}
$$R(r)= -{8a^2\over (1+a^2)^2} {\mu^2 \over (r^2-\mu^2)^2}
\Big({r-\mu\over r+\mu}\Big)^{q}\times$$
{}
$$\times \Big[{1\over B_-\Big({r-\mu\over r+\mu}
\Big)^{2k}+B_+}\Big]^{2\over1+a^2}
\Big[(\delta- k)-{{2kB_-\Big({r-\mu\over r+\mu}\Big)^{2k}}
\over{B_-\Big({r-\mu\over r+\mu}\Big)^{2k}+B_+}}\Big]^2,\eqno(8.2)$$

\noindent where the constants $\delta$ and  $q$
are defined by  $(5.6)$ and $(7.5b)$ respectively. The parameters
$B_\pm$ and $A$ are given as:
{}
$$B_\pm = {1\over2}\Big(1 \pm\sqrt{1 + A^2}\hskip 1pt\Big),
\qquad  A^2=   (1 + a^2){Q^2 e^{2a\phi_0} \over 4\mu^2 m^2}.
\eqno(8.3) $$

The expression (8.2) shows that   the scalar
curvature $R$ is generically divergent on the surface: $r=\mu$.
However, in two special cases  one might get   finite results,
namely: (i) when $q=2$ \footnote{The condition $q=2$ is equvivalent
to the following equation:  $(k-1)^2+{3\over 4}a^2=0$.},
and (ii) when $k=\delta=\pm 1/2$.
As we have seen, the most interesting properties are demonstrated
by the solution $(7.12)$. This solution has one horizon
$\mu_+$ where the metric and both electromagnetic and  scalar fields
are regular. Moreover, it is easy to verify that
the corresponding  scalar curvature $R(r)$ is also regular and
 has the following form:
{}
$$R(r)= -{2a^2\over(1+a^2)^2 }
\Big({r-\mu\over r+\mu}\Big)\times$$
{}
$$\times{(\sqrt{\mu^2+Q_\star^2}-\mu)^2 \over
(r+ \sqrt{\mu^2+Q_\star^2})^4}
\Bigg({r+ \sqrt{\mu^2+Q_\star^2}\over r+\mu}
\Bigg)^{2a^2\over 1+a^2},\eqno(8.4) $$

\noindent where constant $Q_\star$ is defined by the
 expressions $(7.13)$. As it might be expected, $R(r)$ tends to
be zero at the limits $Q\to 0$ and $a\to 0$.  Note that the
scalar curvature $(8.4)$ becomes  zero on
the horizon $r=\mu=\mu_+$ and then changes   sign for $r<\mu_+$.
This suggests that our solution describes the exterior of the
charged black hole only, {\it i.e.} the region outside the
horizon for which $r>\mu_+$. To analyze the behavior of the
solution  (7.12) for distances $r\le \mu_+$, one should
choose  the model  of matter distribution inside the star,
obtain the solution for corresponding field equations in the
 interior region of the star,  and then match both
solutions on the surface. This research is in progress and
will be reported elsewhere.

\section{DISCUSSION.}

 Concluding this paper we would like to present the
final form of the  general static spherically symmetric harmonic
solution
 of the Einstein-Maxwell gravity coupled to the massless scalar
field.  By reconstructing the constant $z_0$  from $(3.7c)$ we will
write this solution in the following parametric form:
{}
$$ ds^2 = {{p^2 -\mu^2}
\over w(p)}dt^2 -{w(p)\over{p^2 -\mu^2}}
r_p^2 dp^2 - w(p)d\Omega, \eqno(9.1)$$
{}
$$ E(p) ={Q r_p\over w(p)} e^{2a\phi(p)}, \eqno(9.2)$$

\noindent where
{}
$$r_p  = {d r\over dp}= 1 +  z_0 \bigg( \ln {{p-\mu}\over
 {p+\mu}} + {2\mu p\over{p^2-\mu^2}}\bigg).$$

\noindent
The  functions $\phi$  and   $w$ from (6.5) and (6.7)
  will be finally:
{}
$$ \phi(p)= \phi_0- {a\over 1 + a^2}\ln\Big[\Big({p+\mu\over
 p-\mu}\Big)^{(k-\delta)} \Big( B_-\Big({p-\mu\over
 p+\mu}\Big)^{2k}+ B_+\Big)\Big],
 \eqno(9.3a) $$
{}
$$ w(p) = ( p^2 - \mu^2)\Big({p+\mu\over p-\mu}\Big)^q
\Big[  B_- \Big({p-\mu\over p+\mu}\Big)^{2k}+B_+
\Big]^{2\over{1+a^2}},  \eqno9.3b)$$
\noindent with the constants $B_\pm$ defined by the expressions (8.3).

Thus, we have presented the class of the solutions (9.3) which
describes the exterior region of the black holes and the naked
singularities and, as any  solution, obtained with the harmonical gauge
condition (2.3), this result has only one
horizon $\mu_+$, which is related to physical mass $\mu_0$ and the
electric charge $Q_0=Qe^{a\phi_0}$ as follows:
{}
$$\mu_+={1\over 4k^2-a^2}\Big[\sqrt{4\mu_0^2k^2-(4k^2-a^2)Q^2_0}-
\mu_0 a\sqrt{1+a^2-4k^2}\Big].\eqno(9.4a) $$
\noindent This expression, independent of the value of the  parameter $k$,
 saturates the usual bound  $\mu_0^2(1+a^2)\ge Q^2_0$. Moreover, the
condition $1+a^2-4k^2\ge0$ suggests that, for an arbitrary values of the
coupling constant $a$, the physically interesting solutions do not
exist if the parameter $k$  belongs to
$|k|>{1\over2}\sqrt{1+a^2}$. Thus, for each
value of the coupling constant $a$, this solution is
characterized by a set of three parameters, the physical mass
$\mu_0$, the electric charge $Q_0$ and the
scalar field parameter $k$. Note that for a spherical body
whose radius is larger than $\mu$, the  constant $k$
generically differs from $\pm 1/2$ and falls into the interval
$|k|\in \Big[0, {1\over2}\sqrt{1+a^2} \Big]$. As we saw, it
takes these values only for black holes or
in the case when a scalar field $\phi$ is totally decoupled
from the matter (besides photons).
The dilaton charge $D$ corresponding to the solution (9.3)
is given as
{}
$$D={1\over 4k^2-a^2}\Big[\sqrt{1+a^2-4k^2}
\sqrt{4\mu_0^2k^2-(4k^2-a^2)Q^2_0}-\mu_0 a\Big].\eqno(9.4b) $$
\noindent Taking $D$ as an independent parameter
of the solution (instead of $Q$), one can present the horizon
(9.4a) as follows:
{}
$$\mu_+={\mu_0a+D\over \sqrt{1+a^2-4k^2}}. \eqno(9.4c)$$

 We have shown that because  this solution was obtained in the
vacuum outside the matter distribution, the result (9.3)
 describes the region which lies outside   the surface (9.4a) only,
{\it i.e.} for $r>\mu_+$.  Although including  a scalar field
in the theory drastically affects the space-time geometry and,
in general, destroys the  horizons, the presence of
arbitrary coupling  constant $a$   gives an opportunity to
explore the behavior of the  results obtained in a different
interaction regimes. Thus, we have found that the solutions for
both gravitational and electromagnetic
fields are not only affected by the scalar field, but also the
non-trivial coupling with matter  constrains the scalar field itself.
Moreover, this opportunity gave us a chance to rule out the non-physical
``inner'' horizons,   which differentiates our results
from those previously obtained in \cite{GM},\cite{GHS}.
Indeed, the quadratic equation for finding the dependence
$\mu=\mu\big(a, k, \mu_0, Q e^{a\phi_0}\big)$ has, in general,
two solutions. It is reasonable to expect that these solutions  $\mu_\pm$
should be regular functions of their arguments.
Moreover, because of the fact that the solution (9.3) describes
the exterior region outside the matter distribution,
 for any values of both parameters $a$ and $k$, the following condition
should be satisfied: $\mu\ge\mu_0\ge0$.
However, it turns out that while the result $\mu_+$ always remains regular,
the solution $\mu_-$, for some physically interesting
values of the parameters $a$ and $k$, might  either be hidden inside the
mass shell $\mu_-<\mu_0$, or it might
be  negative or even divergent\footnote{This analysis exludes one of the
results for $\mu$ in the case of the solution with a negative physical
mass $\mu_0\rightarrow -\mu_0$ as well.}. Such a behavior makes it
possible to rule out the result $\mu_-$  as physically meanningless.

As we discussed in the paper, the presence of an arbitrary
coupling constant $a$ in the Lagrangian (1),  provides an
opportunity to  study the behavior of the solutions in
a different  interaction regimes.
Moreover, we would  like to emphasize here that, in the case of the
interacting fields,  the use of the gauge conditions (2.3)
seems to be more appropriate then the usual Schwarzchild gauge
\cite{GM}-\cite{GHS}. To demonstrate this, let us  analyze the
behavior of the horizons $h_\pm$ corresponding to the solution
\cite{GHS} in the limit of the strong interaction
$a\rightarrow\infty$. Thus, by presenting these horizons in the
equivalent form :
{}
$$h_+=\mu_0+\sqrt{\mu_0+(a^2-1)Q^2}, \eqno(9.5a)$$
{}
$$h_-={a^2+1\over a^2-1}\Big(\sqrt{\mu_0+(a^2-1)Q^2}-\mu_0 \Big),
\eqno(9.5b)$$
\noindent one can find that both expressions above
demonstrate the same tendency when $a\rightarrow\infty$,
namely: $h_+, h_- \rightarrow a Q$. However, this behavior is
in conflict with the model (1) which was used to obtain the solution.
Indeed, as we discussed in the paper, taking the limit
$a\rightarrow \infty$  is equivalent to dropping   both the Maxwell
and the scalar terms from the
expression (1). The remaining Lagrangian should correspond to
that of  general relativity with  the
Shwarzchild solution for a static spherically symmetric case
which has one finite horizon at $h_+=2\mu_0$.
As forour results, that, unlike to these discussed above,
in the limit $a\rightarrow\infty$ from the relation (9.4a) one
obtains $\mu_+\big|_{a\rightarrow \infty} =\mu_0$, which   perfectly
corresponds to the Fock solution (7.4).

Thus, for a the different values of the
coupling constant $a$ we have established
the correspondence of the result obtained (9.3) to a well
known solutions. It should be noted that our solution
naturally contains the harmonical  generalizations of those  discussed
in \cite{GM}, \cite{GHS}. Our initial goal was to find a static,
spherically symmetric, harmonic solution with a regular
geometrical properties, so, the analysis performed
in the  paper, has eliminated many of these  generalizations.
The most interesting result obtained in this paper
is the solution for a charged dilatonic black hole  (7.12) in
harmonical coordinates of the Minkowski  space-time.
This solution   represents a   black hole
with one horizon (and hence for the Kaluza-Klein case),
on which the metric, the scalar
curvature and both electromagnetic and scalar
fields are regular (unlike the solution \cite{GHS}, which has
a regular outer horizon, but those inner horizon is singular).
Because of this property we believe  that
the solution $(7.12)$  will   provide
 an interesting framework for studying different
processes in general relativity as well as in Kaluza-Klein theory.

And, finally, anticipating the possible question: whether or
not the obtained
 solutions are stable, we would like to emphasise  that the stability
of the solution to Einstein-Maxwell
system with an extra free scalar field  Eqs.(7.1) has already
been proven  \cite{ST2}. Also note that  in   \cite{HW}
the stability of the solution  presented
in  \cite{GHS} was inferred for outside the outer horizon.
The special studies of the charged dilatonic black hole solution
Eqs.(7.12) had shown \cite{ST2} that this solution is stable at least
against an axial perturbations. The full analysis of this problem is
currently in progress  and will be reported  in a subsequent publication.

\section{\bf ACKNOWLEDGEDGEMENTS.}

The author wishes to thank Ronald W. Hellings,
Gary T. Horowitz,  Peter K. Silaev and Kip S. Thorne for
valuable and stimulating conversations.
I am also  very grateful to John D. Anderson
for warm hospitality at the JPL.
This work was supported by National Research Council,
Resident Research Associateship award. The research reported in
this publication has been done at the Jet Propulsion Laboratory,
California Institute of Technology, which is
under  contract to the National Aeronautic and Space
Administration.

\end{document}